\documentclass[12pt,preprint]{aastex}
\usepackage{epsfig}

\begin{document}

\title{Modeling Accretion Disk X-ray Continuum of Black Hole Candidates}

\author{Gabor Pszota and Wei Cui\altaffilmark{1}}
\affil{Department of Physics, Purdue University, West Lafayette, IN 47907}
\altaffiltext{1}{Email: pszotag@physics.purdue.edu,cui@physics.purdue.edu}

       
\label{firstpage}

\begin{abstract}
We critically examine issues associated with determining the fundamental 
properties of the black hole and the surrounding accretion disk in an
X-ray binary based on modeling the disk X-ray continuum of the source. We 
base our work mainly on two {\it XMM-Newton} observations of GX 339$-$4, 
because they provided high-quality data at low energies (below 1 keV), 
which are critical for reliably modeling the spectrum of the accretion 
disk. A key issue examined is the determination of the so-called ``color 
correction factor'', which is often empirically introduced to account for 
the deviation of the local disk spectrum from a blackbody (due to electron
scattering). This factor cannot be predetermined theoretically, because 
it may vary 
with, e.g., mass accretion rate, among a number of important factors. 
We follow up on an earlier suggestion to estimate the color correction 
observationally by modeling the disk spectrum with saturated Compton 
scattering. We show that the spectra can be fitted well, and the approach 
yields reasonable values for the color correction factor. For comparison, 
we have also attempted to fit the spectra with other models. We show that 
even the high-soft state continuum (which is dominated by the disk emission) 
cannot be satisfactorily fitted by state-of-the-art disk models. We 
discuss the implications of these results.
\end{abstract}

\section{Introduction}

The X-ray continuum of black hole candidates (BHCs) is roughly composed of two 
main elements (see review by Liang 1998), an ultra-soft component that 
is thought to be associated with emission from the accretion disk, and a hard 
component that is thought to be produced by inverse Compton scattering of 
soft photons by energetic electrons that can be either thermal or non-thermal 
in origin. Modeling the disk component could, in principle, allow one to 
determine the radius of the inner edge of the accretion disk in a BHC (review 
by Tanaka \& Lewin 1995, and references therein). This has been tried and the 
results have provided evidence that the accretion disk extends all the way in 
to the last stable orbit under certain circumstances (Tanaka \& Lewin 1995).
Motivated by this observation, Zhang et al. (1997) suggested that modeling
the X-ray continuum of a BHC could lead to a measurement of the spin of the 
black hole, if the mass of the black hole can be independently derived. 

In retrospect, we now know that the accretion disk reaches the last stable 
orbit probably only in the high-soft state (e.g., Narayan 1996)~\footnote{
It has recently been argued, based on hard-state observations of
BHCs (e.g., Miller et al. 2006), that the disk also reaches the last stable
orbit in the low-hard state. We must, however, caution against drawing
strong conclusions on the properties of the disk from modeling a hard-state
spectrum, because it would require a reliable extraction of the weak disk
component from the dominating hard component whose precise origin (e.g., 
the geometry of the emitting region and the nature of seed photons) is still
being debated (see Cui et al. 2002 for an in-depth discussion). This is why 
we chose to focus 
on the soft-state observations in this work.}, so the
proposed technique may only be applicable to data taken in such a state.
Since the X-ray spectrum of BHCs is dominated by the disk component in the 
high-soft state, the determination of the disk parameters based on spectral
modeling should, in principle, be quite accurate, even if one neglects the
hard component whose physical origin is less well understood, particularly in
the high-soft state. However, there are still serious issues associated with 
the exercise. 

First, the local spectrum of the X-ray emitting portion of the accretion 
disk is not a blackbody, because the opacity is dominated by electron 
scattering.
Saturated Comptonization leads to a ``diluted'' blackbody spectrum, whose 
color temperature is given by $T_{col}=f_{col} T_{eff}$, where $f_{col}$ 
is the color correction factor and $T_{eff}$ is the effective temperature 
(Ebisuzaki et al. 1984). Much effort has gone into finding 
the values of $f_{col}$ that are appropriate for BHCs (Shimura \& Takahara 
1995; Merloni et al. 2000; Davis et al. 2006). The situation is
still uncertain, but it is clear that $f_{col}$ depends on a number of
important physical parameters, such as mass accretion rate, which can 
vary even for a given source. It is, therefore, not possible to know what
value to use {\it a priori}. Cui et al. (2002) proposed an observational
approach to derive $f_{col}$ from the data (see also Shrader \& Titarchuk 
1999). Although the technique showed some promise with limited data, it 
needs to be tested further.

Second, there is observational evidence (Zhang et al. 2000) that the 
surface layer of the accretion disk in BHCs might deviate from the standard 
$\alpha$-disk structure (Shakura \& Sunyaev 1973). Such an effect is
expected from X-ray heating of the disk by a central hard X-ray source (e.g., 
Nayakshin \& Melia 1997; Mistra et al. 1998), but it is not 
clear why the effect is still significant even for the high-soft state, in 
which hard X-ray production is expected to be quite weak. The presence of 
such a ``warm'' layer would add further complication in modeling the 
observed X-ray spectrum (Zhang et al. 2000), because Compton scattering 
in the layer can further modify the spectrum.

Third, some of the widely-used disk models (e.g., the multi-color disk; 
Mitsuda et al. 1984) do not take into account general relativistic effects 
that can affect the formation of the X-ray spectrum. Attempts have been 
made to incorporate the effects empirically in the analysis by introducing 
a number of correction factors (Zhang et al. 1997). Recently, two 
new disk models have been developed that account for the general 
relativistic effects (Li et al. 2005; Davis \& Hubeny 2006). The models also 
consider spectral hardening due to scattering, with one treating $f_{col}$ 
as a free parameter (Li et al. 2005) and the other carrying out radiative 
transfer in the disk (Davis \& Hubeny 2006). The models have been applied to 
observations of a number of BHCs (Shafee et al. 2006; Davis et al. 2006;
McClintock et al. 2006; Middleton et al. 2006).

In this work, we examined some of the issues and also assessed the viability
of the state-of-the-art disk models, making use of data of much 
improved quality that have recently become available. Specifically, we 
analyzed two {\it XMM-Newton} observations of GX 339$-$4 and attempted to 
fit the observed X-ray spectra with different models. With its large 
effective area
and good sensitivity at low energies ($<$ 1 keV), {\it XMM-Newton} offers
distinct advantages over other X-ray observatories for our purposes. The
low-energy sensitivity is often not appreciated as much as it should be; 
it is critical to reliable modeling of the disk spectrum, because the 
effective temperature of the disk is typically $\lesssim$ 1 keV for BHCs.  

\section{Data}
\subsection{XMM-Newton Observations}

We analyzed data from two archival {\it XMM-Newton} observations (ObsIDs 
0093562701 and 0148220201) of GX~339$-$4 during its 2002--2003 outburst. 
The first observation was taken near the peak of the outburst (on 2002 
August 24), judging from 
the ASM/RXTE light
curve~\footnote{See http://heasarc.gsfc.nasa.gov/xte\_weather}, while
the second one was taken at the tail end of the episode (on 2003 March 8). 
GX~339$-$4 was observed for about 61 and 20 ks during the two observations,
respectively.
Since we are mainly interested in the X-ray continuum here, we focused on 
the EPIC data. The pn/EPIC detector was operated in the burst mode, with the 
thin optical blocking filter, during the first observation, and the MOS/EPIC 
detectors were not used. In the second observation, the pn and MOS detectors 
were both run in the timing mode with the medium blocking filter. Even with 
the timing mode, the MOS data still suffer from severe photon pile-up, due to 
the high count rate. In contrast, the pile-up effects are minimal in the pn 
data. This work is, therefore, based on the pn data.

The data were reduced with the standard {\it SAS} package (version 7.0.0).
We followed the procedures described in the {\it XMM-Newton} data analysis
cookbook~\footnote{See http://wave.xray.mpe.mpg.de/xmm/cookbook.} in 
preparing and filtering the data, making light curves, extracting spectra, 
and generating the corresponding arf and rmf files for subsequent spectral 
modeling. We did need to turn off bad-pixel search in processing the first 
observation because of a bug in the searching routine for the burst mode. 
The effects should be negligible because the source was very bright then.
The events of interest were extracted from a rectangular region, with 
RAWX 32--40 RAWY 3--179 and RAWX 34--42 RAWY 3--199 for the 2002 and 2003 
observations, respectively. Filtering expressions ``FLAG = 0'' and ``PATTERN 
$\leq$ 4'' were applied to select good single and double events. 

Because the source was bright during both observations, a significant number 
of source events are present even near the edge of the CCD chip, which makes 
it impossible to cleanly extract background events. This should only affect
the high-energy end of the spectrum (where the background counts may become
comparable or exceed the source counts). Our choice of the central 9 columns 
of the chip was made to minimize the effect on the shape of the spectrum.
However, it led to an underestimation of the overall normalization, which is 
also important here. To determine the normalization more accurately, we also 
made spectra with events from the whole chip. The difference amounts to 
roughly 8\%.
For spectral modeling, we added a 1\% systematic error to the data and grouped 
the channels so that each bin contains at least 500 counts. 

\subsection{RXTE Observations}

To complement the soft-band coverage of {\it XMM-Newton}, we obtained 
simultaneous {\it RXTE} data from the public archive. GX~339$-$4 was
observed with {\it RXTE} for
about 4 and 16 ks, respectively, during the two {\it XMM-Newton} observing 
periods. The data were reduced with {\em FTOOLS 5.2}. We followed the 
standard
steps~\footnote{see http://heasarc.gsfc.nasa.gov/docs/xte/recipes/cook\_book.html}
in preparing and filtering the data, deriving PCA and HEXTE 
spectra from data taken in the standard modes, and generating the 
corresponding response files for spectral modeling. 

A PCA or HEXTE spectrum consists of separate spectra from the individual 
detector units that were in operation. In deriving the PCA 
spectra, we only used data from the first xenon layer of each detector unit 
(which is best calibrated) and combined spectra from all the live detectors 
into one, to maximize the signal-to-noise ratio (S/N). To 
estimate the PCA background, we used the background model for bright 
sources (\mbox{pca\_bkgd\_cmbrightvle\_eMv20030330.mdl}). As for the HEXTE 
data, 
we extracted a spectrum for each of the two clusters separately. 
For spectral modeling, we rebinned the HEXTE spectra so that each bin 
contains at least 5000 counts. We also added a 1\% systematic error to both
the PCA and HEXTE spectra.

\section{Results}

We carried out spectral modeling in {\it XSPEC} (Arnaud 1996). The spectral
bands of interest are 0.5--10 keV (pn/EPIC), 3--25 keV (PCA), and $>$ 15 
keV (HEXTE). The spectra are always jointly fitted with a common model, 
except for a normalization factor (fixed at unity for the pn data) that was 
introduced to account for any residual difference in the calibration of the 
throughput of the detectors. Strictly speaking, however, the {\it XMM-Newton} 
and {\it RXTE} coverages are not always simultaneous, due to the 
difference not only in the observing time but also in the orbits of the two 
satellites. To justify joint modeling, we broke each of the {\it XMM-Newton} 
observations into 8 segments and extracted a spectrum for each segment. We 
compared the individual spectra and observed no apparent variation in the 
shape of the spectrum in either case.

We experimented with several models for the ultra-soft and hard components
of the spectrum. The former is often modeled with a non-relativistic, 
multi-temperature blackbody model (``diskbb'' in XSPEC; Mitsuda et al. 1984). 
For this work, we instead used the two relativistic disk models (``kerrbb'' 
in XSPEC, Li et al. 2005; and ``bhspec'', Davis \& Hubeny 2006). To
test the procedure of deriving the color correction factor from the data, 
as proposed by Cui et al. (2002), we also modeled the disk component with 
saturated Compton scattering (``comptt'' in XSPEC, in a disk geometry; 
Titarchuk 1994). In all cases, the hard component of the spectrum was 
modeled with unsaturated Compton scattering (also ``comptt'' but in a spherical
geometry). Interstellar absorption was taken into account (with ``phabs'' 
in XSPEC). 

The best and only formally acceptable fit to the continuum was obtained 
with \mbox{\it comptt+comptt}. In this case, the residuals reveal the 
presence of discrete features, which include absorption edges at 
863~eV and 880~eV for the 2002 and 2003 observations, respectively, and 
emission lines at 569~eV and 562~eV. We suspect that the edges are 
calibration artifacts, since we were not able to associate them with any
elements. On the other
hand, the emission features could be real, 
with the former being associated with O~VIII and the latter with O~VII 
(corresponding to transitions at rest-frame energies 569 eV and 561 eV,
respectively), which would imply a plasma temperature of 0.1--0.2 keV. 
The lines are unresolved and are quite weak, with equivalent widths of 
26 and 21~eV for the 2002 and 2003 observations, respectively. We will not 
discuss the discrete spectral features any further, since the main focus 
here is on the X-ray continuum. The 2002 
data also show the presence of an emission feature at 2.2~keV, which is 
likely an artifact 
caused by calibration uncertainty around the M-edge of gold (in the mirror 
coating). However, the feature is not apparent in the 2003 data, which is a 
bit puzzling, because the statistics are comparable in the two cases. We 
consulted with the {\it XMM-Newton} Helpdesk about it, and were told
that it had probably been corrected for by the calibration in the
timing mode, but not so well in the burst mode. After accounting for the
discrete spectral features (with ``edge'' and ``gaussian'' in XSPEC), we 
still saw, in the residuals, genuine inconsistency between the pn/EPIC and 
PCA data at low energies, which could be related to known PCA calibration 
uncertainties around the L-edge of xenon. For this work, we resolved the 
issue simply by excluding the PCA data below 9 keV in the joint fits. 

For the 2003 data, the continuum fit also shows significant structures in
the residuals roughly in the range of 5--8 keV, which might be similar to 
those reported by Miller et al. (2004) based on an {\it XMM-Newton}
observation taken several months earlier. They are most likely associated
with the 
K$\alpha$ emission of the iron and its associated absorption edge.
The excess appears broad and asymmetric in shape, as
illustrated in Figure~1. Therefore, we modeled it as a 
gravitationally redshifted disk line (``laor'' in XSPEC; Laor 1991). Also, 
we included a smeared edge (``smedge'' in XSPEC) in the fit. The results are: 
$E_{Laor}=6.48^{+0.07}_{-0.09}$ keV, $i=51$\arcdeg\,$^{+2}_{-1}$, 
$q = 5.2\pm 0.2$, and $R_{in} = 1.76^{+0.10}_{-0.06}$ $R_g$ 
(where $R_g$ is the gravitational radius) for the line; 
$E_{edge} = 8.5\pm 0.1$ keV, $W = 2.7^{+0.5}_{-0.4}$ keV, and 
$\tau=0.59^{+0.07}_{-0.05}$ for the edge. Note that we fixed $R_{out}$ at 
$400$ $R_g$ in the ``laor'' model. The obtained value for the 
inclination angle ($i$) is consistent with those estimated for the 
system (e.g., Zdziarski et al. 2004). If this interpretation is correct, 
the results would require a very high value ($a^* \gtrsim 0.97$) for the 
black hole spin (cf. Miller et al. 2004). However, no such broad line 
(nor the edge) is apparent in the 2002 data. Adding the line (as a 
Gaussian component) to the model, we found that the data could accommodate 
it, but its equivalent width would be merely $14^{+12}_{-9}$ eV, compared 
to $485^{+217}_{-130}$ eV based on the 2003 data.  
 
Figure~2 shows the observed X-ray spectra of GX 339$-$4, along with the 
best-fit models and the associated residuals. The parameters of the 
continuum fits are summarized in Table~1. The source was clearly in the
high-soft state during the 2002 observation, with the disk contributing
about 96\% of the 0.5--10 keV flux. The spectrum became harder during 
the 2003 observation, but the disk still contributed about 80\% of the 
0.5--10 keV flux. Following Cui et al. (2002), we attempted to derive
the color correction factor from the continuum fits. Briefly, to account
for the effects of scattering in a Shakura-Sunyaev disk
(Shakura \& Sunyaev 1973), one should, strictly speaking, start with a
multitemperature blackbody spectrum for the seed photons.
However, {\it comptt} assumes a Wien spectrum for the seed photons.
Fitting the peak of {\it diskbb} with a Wien distribution leads
to $T_{diskbb} = 2.7T_{Wien}$. Based on spectral modeling
with {\it comptt}, therefore, we can approximate the color correction factor
as $f_{col} = T_e/2.7T_0$
(Cui et al. 2002; see also Zhang 2005). For the 2002 and 2003 
observations, respectively, we have
$f_{col} = 1.48^{+0.09}_{-0.08}$ and 
$1.35^{+0.01}_{-0.01}$, which seem quite reasonable. This lends support to
the viability of the observational approach in deriving $f_{col}$.

We then replaced the saturated Compton component with a multicolor disk 
model, but failed to obtain any formally acceptable fits to the observed 
X-ray continua with either ``kerrbb'' or ``bhspec''. In this case, 
we fixed the inclination angle at the value from relativistic line modeling 
(51\arcdeg), the mass of the black hole at $10~M_{\sun}$, and the distance 
at 8 kpc (Zdziarski et al. 2004). With ``kerrbb'', we also adopted the 
default settings for torque-free inner boundary condition, returning 
radiation, and limb darkening, and fixed the normalization at unity and 
the color correction factors at the values that we derived.
The best-fit models are shown in Figure~3.
Neither one is formally acceptable, with $\chi^2$/dof = 2634/1203 and 
2010/1079 for the 2002 and 2003 observations, respectively. The residuals
show significant structures in both cases. Taken at its face value, the
black hole spin would be about 0.7, after correcting for the loss of flux 
due to the use of the central nine columns of the pn chip (see \S~2.1). 
The situation is hardly improved when the inclination angle and the
color correction factor are allowed to vary.

Figure~4 shows the best-fit models with ``bhspec''.
Again, significant features are noticeable in
the residuals. The $\chi^2$ values of the fits are $\chi^2$/dof = 2246/1203 
and 2505/1079 for the 2002 and 2003 observations, respectively. As already 
mentioned, in this model spectral hardening (due to electron scattering) 
is taken into account in modeling the disk atmosphere. Again, taken at 
its face value, the black hole spin is about 0.5. Relaxing the inclination 
angle does not improve the fits. 

\section{Discussion}

The importance of accurately modeling the accretion disk X-ray continuum of
BHCs goes beyond gaining insights into radiative processes associated with
accretion flows. It also lies in the exciting prospect of deriving the spin 
of black holes from such spectral modeling. The technique is one of many 
that have been proposed for BHCs (Laor 1991; Bromley et al. 1997; Zhang at al.
1997; Nowak et al. 1997; Cui et al. 1998; 
Stella et al. 1999; Wagoner et al. 2001; Abramowicz \& Kluzniak 2001).
Although varying
degrees of success have been achieved, it is fair to say that the techniques
all have serious issues in their applications to the data. Further 
investigation, both theoretical and observational, is thus needed to examine 
the issues. 

We have demonstrated in this work that the high quality of the data is 
starting to demand a proper treatment of electron scattering in radiative 
transfer through the accretion disk around a stellar-mass black hole. 
Some of the effects that were not appreciated previously in fitting low 
S/N data are now becoming apparent. At present, this
demanding situation fundamentally limits our ability to reliably derive 
the physical parameters of the accretion disk or the black hole in an 
X-ray binary, based on modeling the disk X-ray continuum. There are also
observational issues that add additional uncertainties to the exercise. 
For instance, many key parameters (e.g., black hole mass, inclination 
angle, and distance) that characterize a source are often poorly 
determined but are needed to determine, e.g.,  the black hole spin. 
This is entirely independent of the quality of X-ray data. Also, perhaps 
less appreciated are the significant uncertainties in the absolute and 
cross calibrations of the detectors on different X-ray satellites. This 
issue is relevant, because the determination of the spin of a black hole 
in an X-ray binary depends critically on the overall normalization of 
the X-ray continuum. This is the reason why one must be very careful in 
comparing results based on data from different satellites.

We have shown that neither of the two state-of-the-art disk models is 
capable of satisfactorily fitting the observed ultra-soft component of 
the spectra of GX 339$-$4.
While this is perhaps not totally surprising for ``kerrbb'', since it
does not actually carry out radiative transfer calculations, it is for
``bhspec''. These models have recently been applied to data to derive
the spin of black holes in a number of systems, so our finding is 
somewhat disappointing. If we take the best-fit parameters at their 
face values, the models would suggest that 
GX~339$-$4 contains a moderately rotating black hole (with $a^* \sim$ 
0.5--0.6). On the other hand, if we attribute the asymmetry in the 
profile of the observed Fe K$_{\alpha}$ line to gravitational redshift, 
we would conclude that the source contains a rapidly rotating black 
hole (with $a^* \approx 0.96$). We should note, however, that the apparent
inconsistency can be easily reconciled when we take into account the 
large uncertainties
associated with, e.g., black hole mass, inclination angle, and distance.
For example, if we adopt $13.5~M_{\sun}$ for the black hole mass, 
51\arcdeg\ for the inclination, and $7.5$~kpc for the distance, the 
``kerrbb' model yields $a^* \approx 0.93$ and $0.96$ when fitting the
2002 and 2003 data, respectively.

We were able to fit the ultrasoft component quite satisfactorily with 
a simple saturated Compton scattering model. The results allowed us to 
test a procedure that was previously suggested by Cui et al. (2002) to 
empirically derive the color correction factor from the same X-ray data.
The values obtained are very close to the theoretical expectation (e.g.,
Shimura \& Takahara 1995), which is also often adopted in spectral 
modeling. Therefore, our results have provided further support for 
this observational approach. Although the use of a single color 
correction factor ignores possible radial dependence of spectral
hardening in the disk, it does not seem unreasonable given that the 
X-ray emission from the disk originates from a relatively narrow 
region (closest to the black hole).

\section{Conclusions}

Based on our joint spectral analysis of two simultaneous XMM-Newton/RXTE 
observations of GX 339-4, we can draw following conclusions:
\begin{itemize}
\item The empirical procedure to derive the color correction
factor ($f_{col}$) observationally, as proposed by Cui et al. (2002), yields
reasonable results. If confirmed by further investigations, this would 
eliminate a major (theoretical) uncertainty in deriving the parameters of 
the disk from modeling the X-ray continuum.
\item The observed X-ray continuum of GX 339-4 in the high-soft state, which
is dominated by emission from the optically-thick accretion disk, cannot
be satisfactorily fitted by any existing disk model. Therefore, one should
excise caution in assessing quantitative results from such spectral modeling.
\end{itemize}

\begin{acknowledgements}

We wish to thank Shuangnan Zhang for suggesting the derivation of the 
spectral hardening factor 
from modeling the disk X-ray continuum and for subsequently collaborating on
the subject. This work is a follow-up to much of the initial discussions. We
also thank Lev Titarchuk for candid discussions on the theoretical aspects 
of the subject.
This research has made use of data obtained through the High Energy 
Astrophysics Science Archive Research Center Online Service, provided by 
the NASA/Goddard Space Flight Center. It was supported in part by NASA 
through the LTSA grant NAG5-9998. We also gratefully acknowledge financial
support from the Purdue Research Foundation and from a Grodzins Summer 
Research Award from the Department of Physics at Purdue University (to G.P.). 

\end{acknowledgements}

\clearpage

\begin{deluxetable}{lcccccccccc}
\tabletypesize{\scriptsize}
\tablecolumns{11}
\tablewidth{0pc}
\rotate
\tablecaption{Best X-ray Continuum Fits\tablenotemark{a}}
\tablehead{
 & & \multicolumn{4}{c}{comptt} & \multicolumn{4}{c}{comptt} &  \\
\cline{3-6} \cline{7-10} \\
\colhead{Obs} & \colhead{$N_H$} & \colhead{$kT_0$} & \colhead{$kT_e$} & \colhead{$\tau$} & \colhead{$K$} & \colhead{$kT_0$} & \colhead{$kT_e$} & \colhead{$\tau$} & \colhead{$K$} & \colhead{$\chi^2$/dof} \\
 & $10^{21}\mbox{ }cm^{-2}$ & keV & keV & & & keV & keV & & & 
}
\startdata
2002 & $4.5$ $(+1 -2)$ & $0.20$ $(1)$ & $0.793$ $(+3 -4)$ & $13.4$ $(2)$ & $25$ $(+2 -1)$ & $1.7$ $(+2 -1)$ & $46^{+56}_{-21}$ & $1.8$ $(+1 -2)$ & $1.7^{+1.7}_{-1.4}\times 10^{-3}$ & 978/1201 \\
2003 & $4.75$ $(1)$ & $0.170$ $(1)$ & $0.618$ $(1)$ & $10.07$ $(2)$ & $7.58$ $(2)$ & $1.11$ $(1)$ & $183$ $(2)$ & $0.38$ $(2)$ & $3.21$ $(3)\times 10^{-3}$ & 920/1076 \\ \hline
\enddata
\tablenotetext{a}{The numbers in parentheses indicate uncertainty in the last 
digit. For asymmetric errors, both the lower and upper bounds are shown, 
again for the last digit. The errors shown represent 90\% confidence 
intervals for single parameter estimation. }
\end{deluxetable}

\clearpage

\begin{figure}
\includegraphics[width=4.5in,angle=-90]{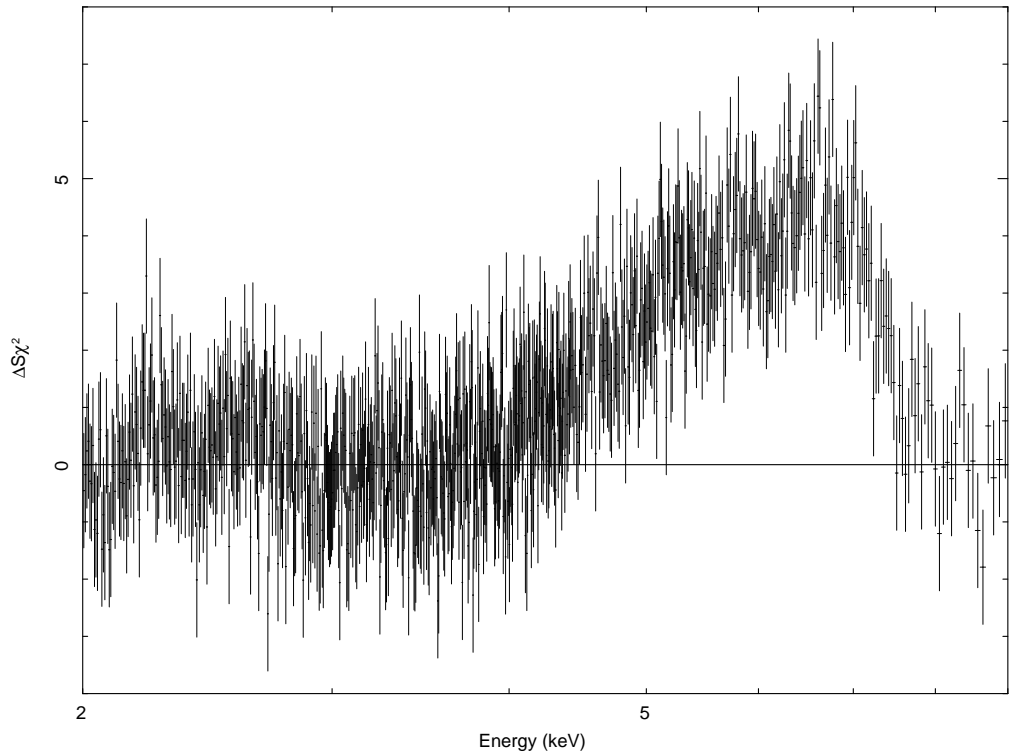}
\caption{Broad line detected in the 2003 X-ray spectrum. Shown are the
residuals after the ``laor'' component is removed from the best-fit model
(see text). }
\end{figure}

\begin{figure}
\includegraphics[width=2.5in,angle=-90]{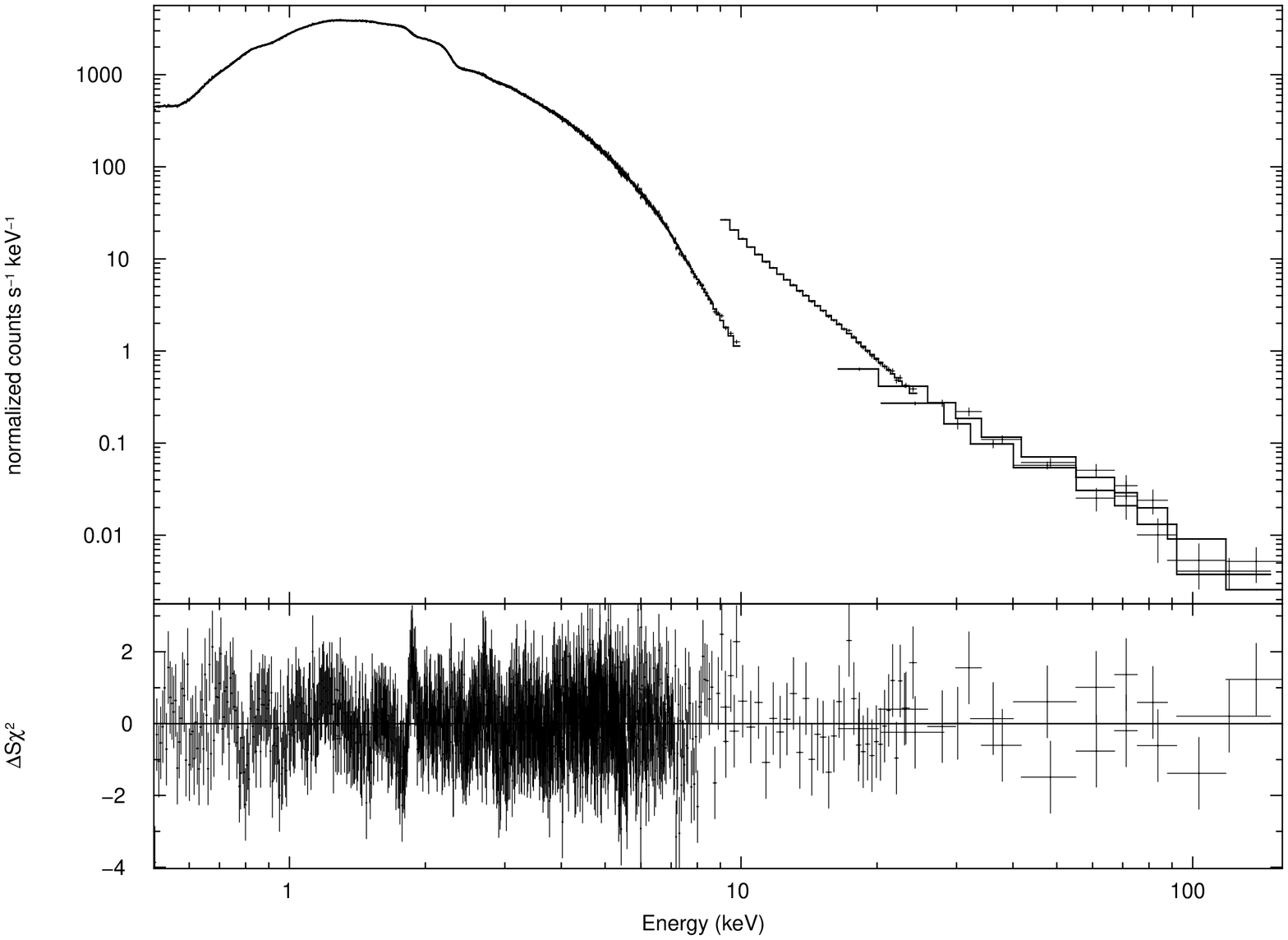}
\includegraphics[width=2.5in,angle=-90]{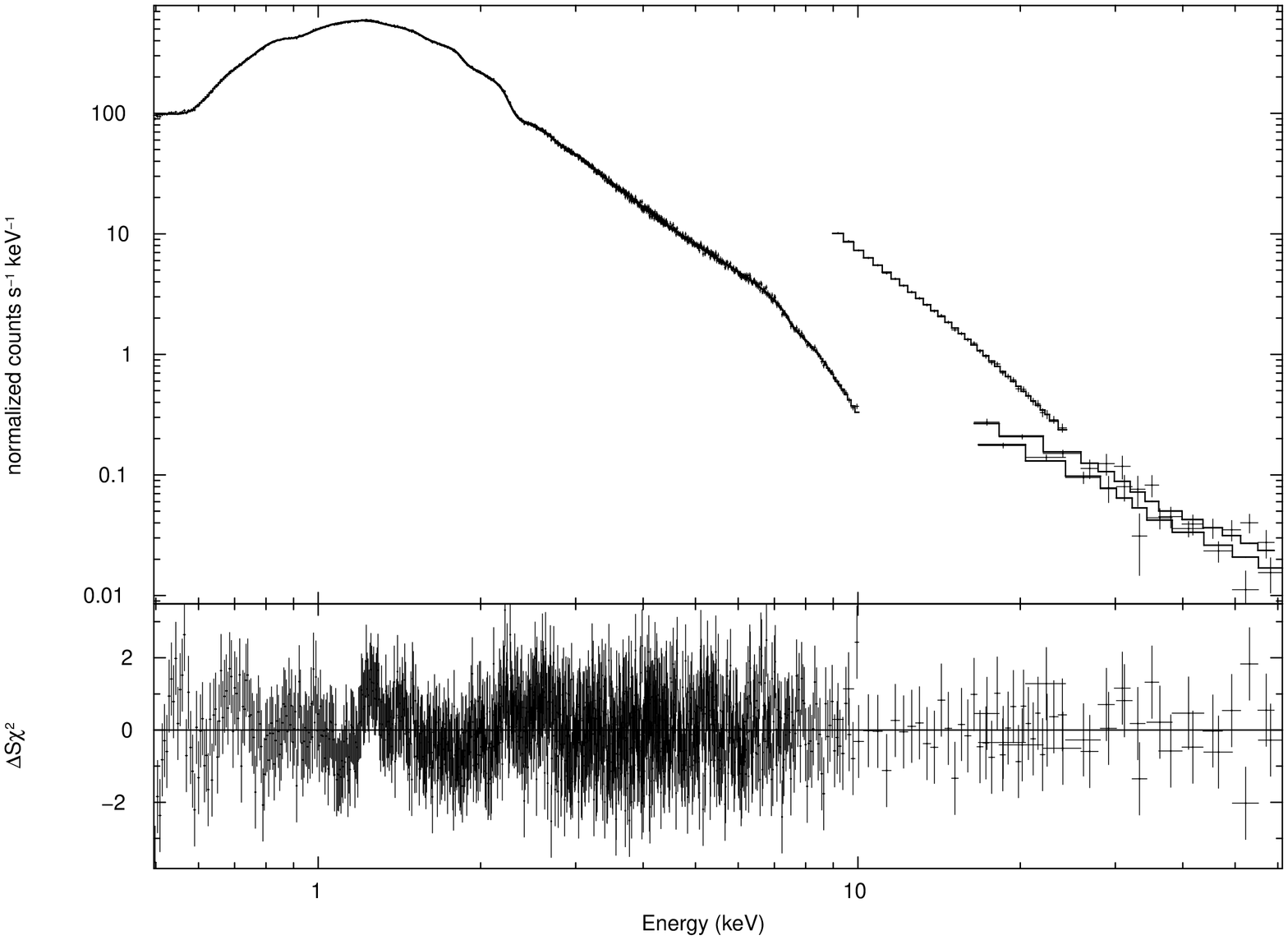}
\caption{Observed X-ray spectra of GX~339$-$4 from the 2002 ({\it left}) and 
2003 ({\it right}) observations. The best-fit models are shown in solid 
histograms. The bottom panels show the respective residuals of the fits. }
\end{figure}

\begin{figure}
\includegraphics[width=2.5in,angle=-90]{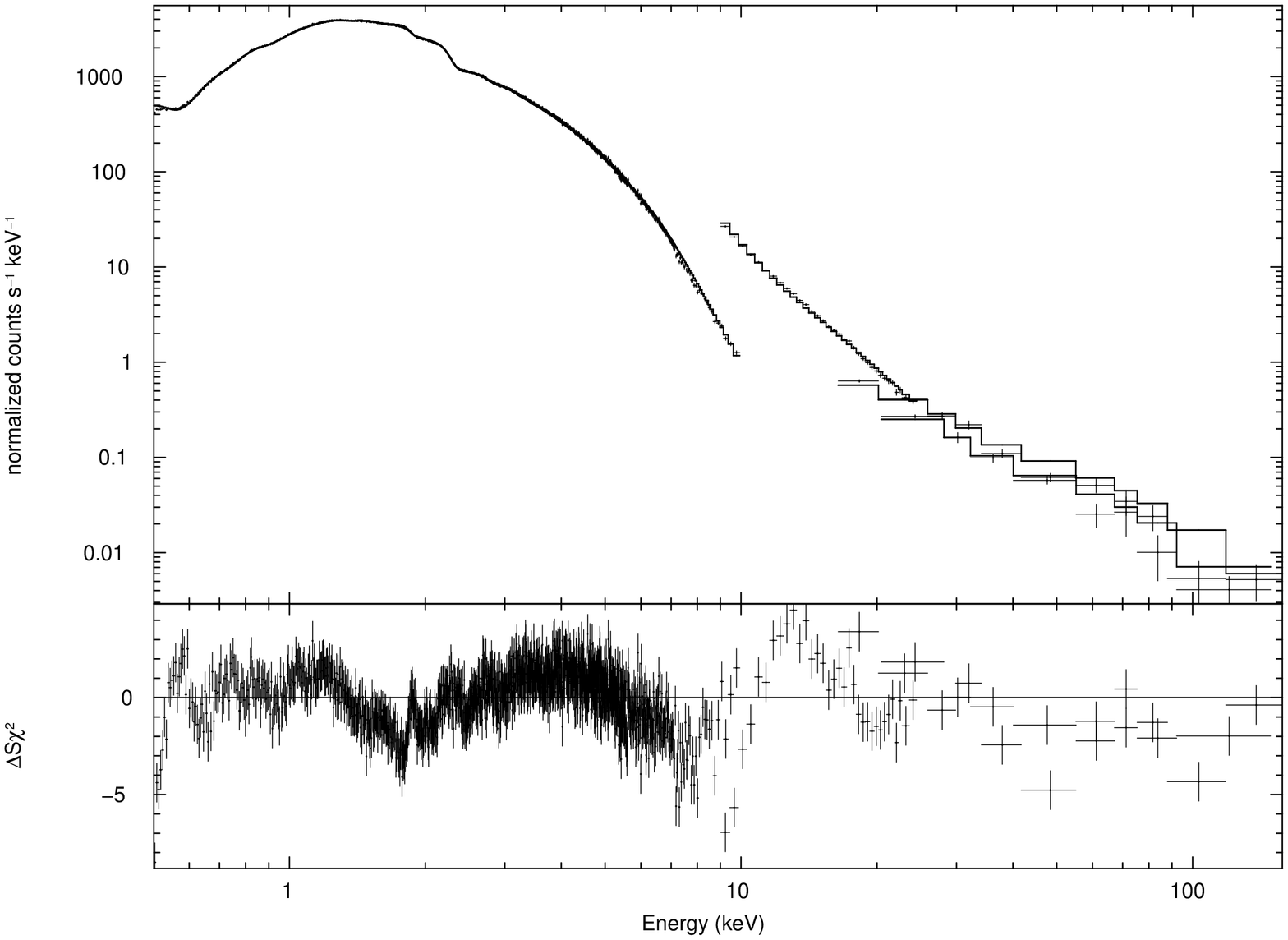}
\includegraphics[width=2.5in,angle=-90]{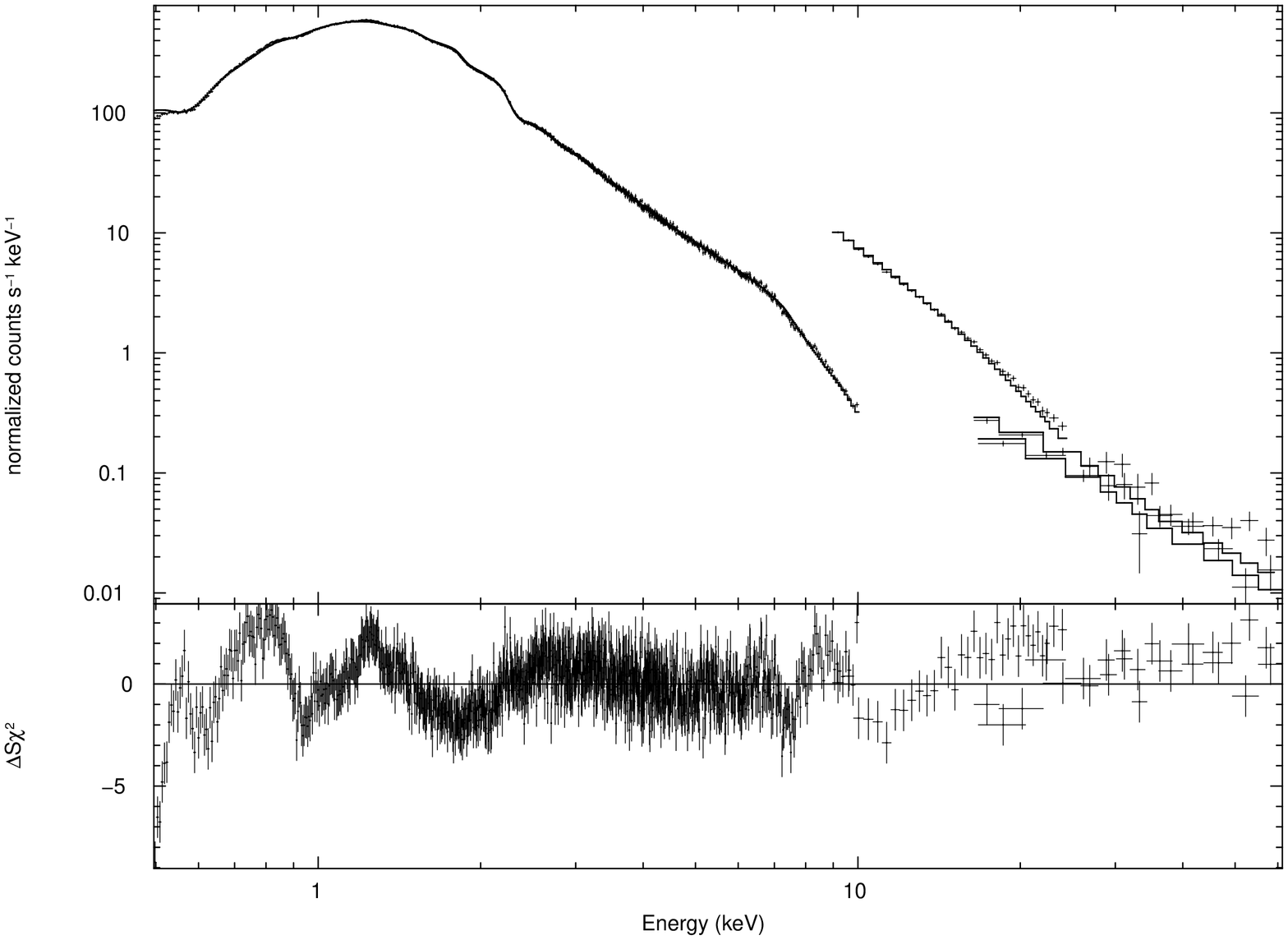}
\caption{Same as Fig.~2 but the disk emission was modeled with ``kerrbb''. }
\end{figure}

\begin{figure}
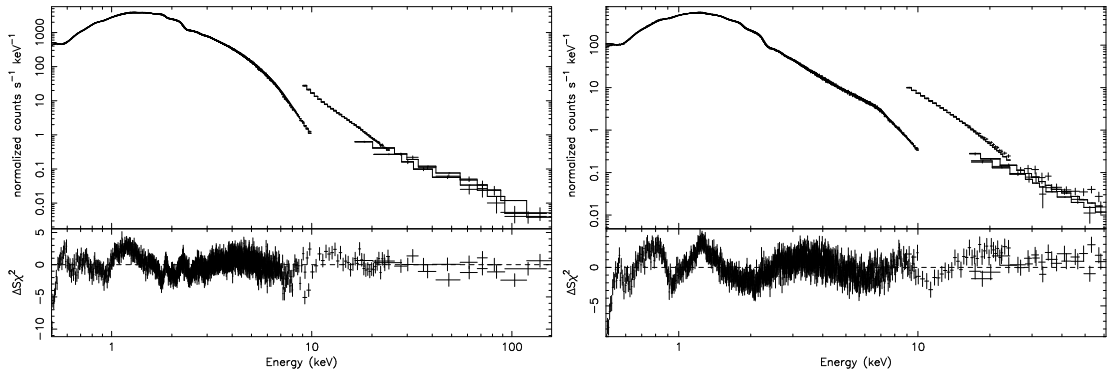

\includegraphics[width=1.9in,angle=-90]{f4a_bw.eps}
\includegraphics[width=1.9in,angle=-90]{f4b_bw.eps}
\caption{Same as Fig.~2 but the disk emission was modeled with ``bhspec''.}
\end{figure}

\end{document}